\documentclass[twocolumn,showpacs,superscriptaddress,amsmath,amssymb,times,pre]{revtex4}

\usepackage{graphicx}% Include figure files
\usepackage{dcolumn}% Align table columns on decimal point
\usepackage{bm}% bold math
\usepackage{amsmath}

\begin{document}

\title{Greedy routing on networks of mobile agents}

\author{Han-Xin Yang}
 \affiliation{Department of Physics, Fuzhou University, Fuzhou 350002, China}

\author{Wen-Xu Wang}
\affiliation{School of Electrical, Computer and Energy Engineering,
Arizona State University, Tempe, AZ 85287, USA}

\author{Ying-Cheng Lai}
\affiliation{School of Electrical, Computer and Energy Engineering,
Arizona State University, Tempe, AZ 85287, USA}

\author{Bing-Hong Wang}
\affiliation{Department of Modern Physics, University of Science and
Technology of China, Hefei 230026, China}

\begin{abstract}

In this paper, we design a greedy routing on networks of mobile
agents. In the greedy routing algorithm, every time step a packet in
agent $i$ is delivered to the agent $j$ whose distance from the
destination is shortest among searched neighbors of agent $i$. Based
on the greedy routing, we study the traffic dynamics and
traffic-driven epidemic spreading on networks of mobile agents. We
find that the transportation capacity of networks and the epidemic
threshold increase as the communication radius increases. For
moderate moving speed, the transportation capacity of networks is
the highest and the epidemic threshold maintains a large value.
These results can help controlling the traffic congestion and
epidemic spreading on mobile networks.

\end{abstract}

\date{\today}


\maketitle
\section{Introduction}\label{sec:intro}
Due to the increasing importance of communication networks such as
the Internet~\cite{Internet} and networks of mobile phone
users~\cite{mobile phone} in modern society, the traffic of
information flows~\cite{Traffic_Dynamics} and the spreading of
computer viruses~\cite{Epidemic_Spreading} in these networks have
attracted more and more attention.

Traffic dynamics concern mainly how to effectively deliver
information packets from source to destination on a network and
avoid traffic congestion. The study of epidemic spreading focuses on
the forecast and control of computer viruses. For a long time, the
two types of dynamical processes were investigated separately.
Recently, Meloni et al. \cite{PNAS} introduced a theoretical
approach to incorporate traffic dynamics in virus spreading. In
particular, they cast the susceptible-infected-susceptible (SIS)
model \cite{SIS} in a flow scenario where contagion is carried by
information packets traveling across the network. A susceptible node
is more likely to be infected if it receives more packets from
infected neighbors. They found that the epidemic threshold depends
on the traffic and decreases as traffic flow increases.

In communication networks, information packets are forwarded from
sources to destinations by specific routing protocol. To enhance the
the transportation capacity of communication networks, researchers
have designed various routing algorithms, including the shortest
path~\cite{shortest1,shortest2}, the integration of static and
dynamic information~\cite{integrate}, the local
routing~\cite{local}, the efficient routing~\cite{efficient}, the
greedy algorithm~\cite{greedy}, and so on. It has been found that
the routing algorithm also plays an important role in traffic-driven
epidemic spreading~\cite{routing1,routing2}.

Previous studies about traffic dynamics usually focus on static
networks, where nodes are motionless and links among nodes keep
fixed. In our recent paper~\cite{yang}, we have studied the
transportation dynamics on networks of mobile agents. We assume that
agents move on a plane and the searching area of a agent $i$ is a
circle centered at $i$. In that paper, information packets were
delivered according to random routing algorithm, that is, a packet
in agent $i$ is forwarded to a randomly chosen agent $j$ from agent
$i$'s searching area. The random routing can be applied in the case
where moving agents cannot obtain the information about other
agents' positions. However, if an agent can know other agents'
positions, the random routing is not an effective algorithm for a
packet to quickly reach the destination. Utilizing the information
of agents' positions, we now propose a greedy routing on networks of
mobile agents. In the greedy routing, every time step a packet in
agent $i$ is delivered to the agent $j$ whose distance from the
destination is shortest among agent $i$'s searched neighbors. We
have found that the greedy routing markedly enhances the
transportation capacity of networks compared with random routing.
Based on the greedy routing, we study how moving speed and
communication radius affects traffic dynamics and traffic-driven
epidemic spreading.

The paper is organized as follows. In Sec. \ref{sec:model}, we
describe our model in terms of the greedy routing and epidemic
spreading. Results on the traffic dynamics and the epidemic
spreading are presented in Sec. \ref{sec:traffic} and Sec.
\ref{sec:epidemic} respectively. A brief conclusion is given in Sec.
\ref{sec:conclusion}.

\section{Model}\label{sec:model}

In our model, $N$ agents move on a square-shaped cell of size $L$
with periodic boundary conditions. Agents change their directions of
motion $\theta$ as time evolves, but the moving speed $v$ is the
same for all agents. Initially, agents are randomly distributed on
the cell. After each time step, the position and moving direction of
an arbitrary agent $i$ are updated according to
\begin{equation}
x_{i}(t+1)=x_{i}(t)+v\cos\theta_{i}(t),
 \label{1}
\end{equation}
\begin{equation}
y_{i}(t+1)=y_{i}(t)+v\sin\theta_{i}(t),
 \label{2}
\end{equation}
\begin{equation}
\theta_{i}(t)=\Psi_{i},
 \label{3}
\end{equation}
where $x_{i}(t)$ and $y_{i}(t)$ are the coordinates of the agent at
time $t$, and $\Psi_{i}$ is an $N$-independent random variable
uniformly distributed in the interval $[-\pi,\pi]$.

Each agent has the same communication radius $r$. Two agents can
communicate with each other if the distance between them is less
than $r$. We define an agent $i$'s neighbors as agents who are
within agent $i$'s communication area. At each time step, there are
$R$ packets generated in the system, with randomly chosen source and
destination agents, and each agent can deliver at most $C$ packet
toward its destination. To transport a packet whose destination, an
agent performs a local search within a circle of radius $r$. If the
packet's destination is found within the searched area, it will be
delivered directly to the destination. Otherwise, the packet is
forwarded to a according to the greedy routing, which is defined as
follows.

At time $t$, agent $i$ has a packet whose destination is agent $j$.
Agent $k$ is one of $i$'s neighbors at that moment. The distance
between agent $k$ and agent $j$ is
\begin{equation}
d_{jk}(t)=\sqrt{[x_{j}(t)-x_{k}(t)]^{2}+[y_{j}(t)-y_{k}(t)]^{2}}.
 \label{4}
\end{equation}
Agent $i$ will deliver the packet to the agent $k$ whose distance
$d_{jk}(t)$ is shortest among agent $i$'s neighbors.

The queue length of each agent is assumed to be unlimited and the
first-in-first-out principle holds for the queue. Once a packet
reaches its destination, it will be removed from the system.¡£

After a transient time, the total number of delivered packets at
each time will reach to a steady value, then an initial fraction of
agents $\rho_{0}$ is set to be infected (e.g., we set $\rho_{0}=0.1$
in numerical experiments). The infection spreads in the network
through packet exchanges. For example, at time $t$ agent $i$ is
infected and a packet is traveling from agent $i$ to a susceptible
agent $j$, then at the next time step, agent $j$ will be infected
with probability $\beta$. The infected agents are recovered at rate
$\mu$ (we set $\mu=1$ in this paper).

\section{Traffic dynamics}\label{sec:traffic}

In this section, we study traffic dynamics based on the greedy
routing. We set the number of agents $N=1500$, the size of the
square region $L = 10$ and the delivering ability of each agent
$C=1$ in this section.

To characterize the transportation capacity of a network, we exploit
the order parameter $\eta$ introduced in Ref.~\cite{order}:
\begin{equation}
\eta(R)=\lim_{t\rightarrow\infty} \frac{C}{R}\frac{\langle\Delta
N_{p}\rangle}{\Delta t},
 \label{5}
\end{equation}
where $\Delta N_{p}=N_{p}(t+\Delta t)-N_{p}(t)$, $\langle \cdot
\cdot \cdot \rangle$ indicates the average over a time window of
width $\Delta t$, and $N_{p}(t)$ represents the total number of
information packets in the whole network at time $t$.

As the packet-generation rate $R$ is increased through a critical
value of $R_{c}$, a transition occurs from free flow to congestion.
For $R \leq R_{c}$, there is a balance between the number of
generated and that of removed packets so that $\langle \Delta
N_{p}\rangle= 0$, leading to $\eta(R)= 0$. In contrast, for
$R>R_{c}$, congestion occurs and packets will accumulate in the
system, resulting in a positive value of $\eta(R)$. The
transportation capacity of a network can thus be characterized by
the critical value $R_c$.

\begin{figure}
\begin{center}
\scalebox{0.39}[0.39]{\includegraphics{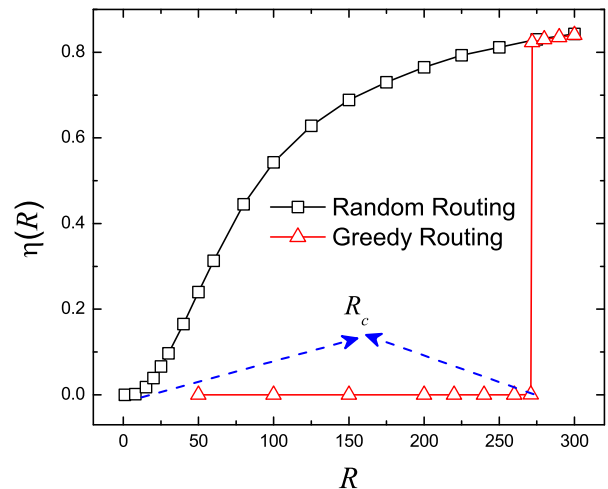}} \caption{(Color
online) The order parameter $\eta(R)$ as a function of the
packet-generation rate $R$ for random routing and greedy routing.
For both routing algorithms, the moving speed $v=0.1$ and the
communication radius $r=1$. $\eta(R)$ is obtained by averaging over
$5\times 10^{4}$ time steps after disregarding $5\times 10^{3}$
initial steps as transients. Each data point results from an average
over 50 different realizations.} \label{fig1}
\end{center}
\end{figure}

Figure~\ref{fig1} shows that the order parameter $\eta(R)$ as a
function of $R$ for random routing and greedy routing. One can see
that the critical value $R_c$ for greedy routing is much larger than
that of random routing when other parameters are the same. For the
moving speed $v=0.1$ and the communication radius $r=1$, the
critical value $R_c$ for random routing and greedy routing is about
8 and 270 respectively. This result shows that the greedy routing
can greatly enhance the transportation capacity of networks in
comparison with that of the random routing. In the following, we
will detailedly study the traffic dynamics based on the greedy
routing.

\begin{figure}
\begin{center}
\scalebox{0.41}[0.41]{\includegraphics{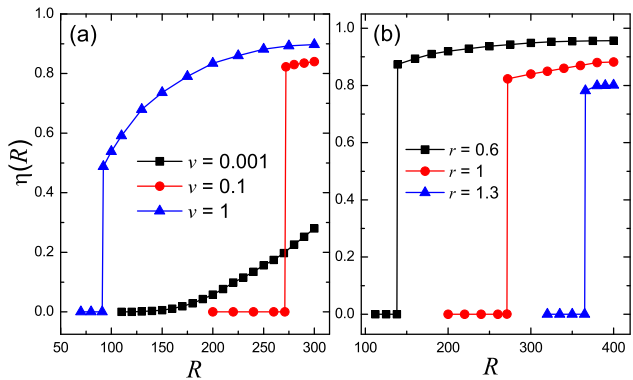}} \caption{(Color
online) (a) The order parameter $\eta(R)$ as a function of the
packet-generation rate $R$ for different values of the moving speed
$v$. The communication radius $r=1$. (b) The order parameter
$\eta(R)$ vs $R$ for different values of $r$. The moving speed
$v=0.1$. Each data point results from an average over 50 different
realizations.} \label{fig2}
\end{center}
\end{figure}

\begin{figure}
\begin{center}
\scalebox{0.4}[0.4]{\includegraphics{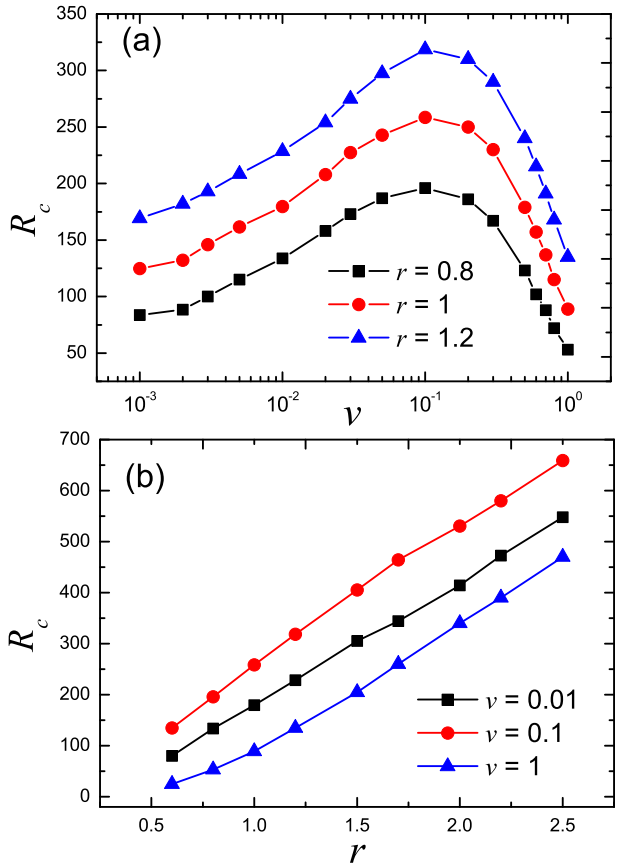}} \caption{(Color
online) (a) The critical value $R_{c}$ as a function of the moving
speed $v$ for different values of the communication radius $r$. (b)
The dependence of $R_{c}$ on $r$ for different values of $v$. Each
data point results from an average over 50 different realizations. }
\label{fig3}
\end{center}
\end{figure}

\begin{figure}
\begin{center}
\scalebox{0.38}[0.38]{\includegraphics{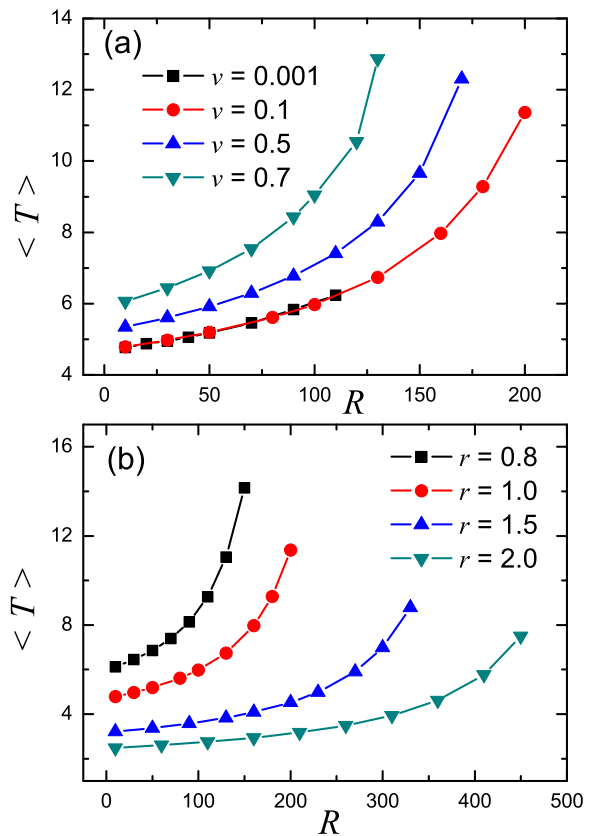}} \caption{(Color
online) (a) The average traveling time $\langle T \rangle$ as a
function of the packet-generation rate $R$ ($<R_{c}$) for different
values of the moving speed $v$. The communication radius $r=1$. (b)
The dependence of $\langle T \rangle$ on $R$ for different values of
$r$. The moving speed $v=0.1$.} \label{fig4}
\end{center}
\end{figure}

\begin{figure*}
\begin{center}
\scalebox{0.87}[0.87]{\includegraphics{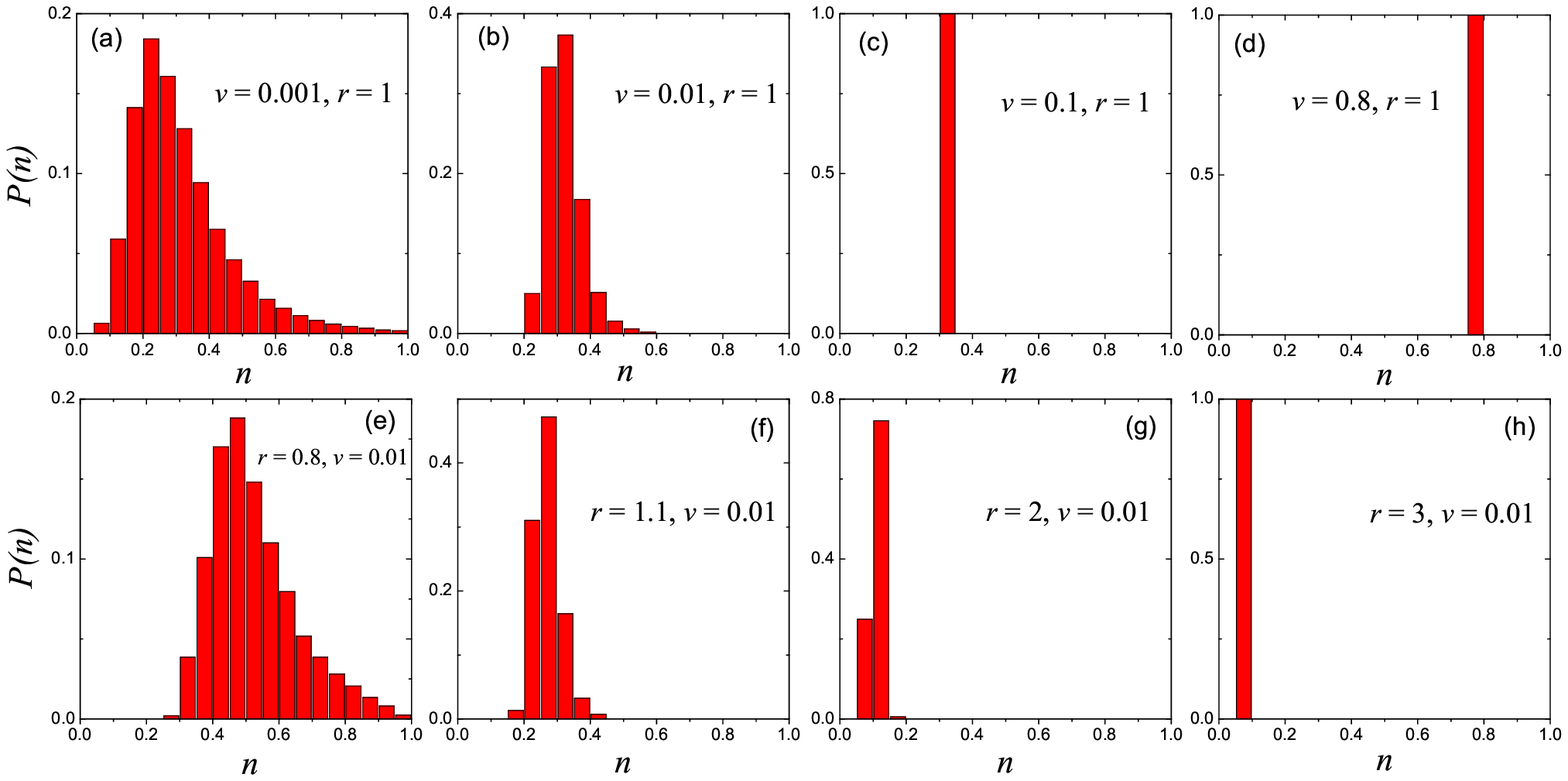}} \caption{(Color
online) The load distribution $P(n)$ vs $n$ for different values of
the moving speed $v$ and the communication radius $r$. The
packet-generation rate $R=100$. From (a) to (d), $v$ is 0.001, 0.01,
0.1, 0.8 respectively and $r$ is fixed to be 1. From (e) to (h), $r$
is 0.8, 1,1, 2, 3 respectively and $v$ is fixed to be 0.01.
 } \label{fig5}
\end{center}
\end{figure*}

Figure~\ref{fig2} shows that the order parameter $\eta(R)$ as a
function of $R$ for different values of $v$ and $r$. One can see
that both the moving speed and the communication radius affect the
onset of traffic congestion. It is interesting to find that there
are two different kinds of phase transition from free flow to
congestion. When the moving speed is very small (i.e., $v=0.001$),
the jamming transition is a second-order phase transition. When the
moving speed is large enough (i.e., $v=0.1$ and $v=1$), the
appearance of a congested phase belongs to a first-order phase
transition.

Since the transportation capacity of networks is characterized by
the critical value $R_{c}$, it is of particular interest for us to
study how $R_{c}$ is affected by the moving speed $v$ and the
communication radius $r$. Figure~\ref{fig3}(a) shows the dependence
of $R_{c}$ on $v$ for different values of $r$. One can observe a
nonmonotonic behavior. For a fixed value of $r$, the largest value
of $R_{c}$ is obtained when the moving speed $v$ is moderate (about
0.1). This result is different from that of Ref.~\cite{yang}, where
$R_{c}$ increases as $v$ increases for random routing.
Figure~\ref{fig3}(b) shows the dependence of $R_{c}$ on $r$ for
different values of $v$. One can see that, $R_{c}$ increases as the
communication radius $r$ increases regardless of the values of $v$.

To understand how the moving speed and communication radius affect
the transportation capacity of networks, we study the average
traveling time $\langle T \rangle$ and the load distribution $P(n)$
in the next.

The average traveling time $\langle T \rangle$ is defined as the
average time steps for a packet traveling from source to
destination. We investigate the average traveling time $\langle T
\rangle$ as a function of the packet-generation rate $R$ in the free
flow state ($R\leq R_{c}$). From Fig.~\ref{fig4}, one can find that
$\langle T \rangle$ increases as $R$ increases. Due to the
randomness of movement and packet generation, the number of packets
in an agent's queue fluctuates as time evolves. At a time, if the
number of packets in an agent's queue excesses the agent's
delivering ability, extra packets must wait for being delivered in
the next time step. As $R$ increases, the above phenomenon of
transient congestion occurs more frequently, leading to a longer
$\langle T \rangle$. For a fixed value of $R$, we find that $\langle
T \rangle$ keeps almost unchanged when the moving speed $v$
increases from 0.001 to 0.1 [see Fig.~\ref{fig4}(a)]. But as $v$
continually increases, $\langle T \rangle$ increases. As is shown in
Fig.~\ref{fig4}(a), for the same $R$, the average traveling time
$\langle T \rangle$ increases as $v$ increases from 0.1 to 0.7.
Figure~\ref{fig4}(b) shows the average traveling time $\langle T
\rangle$ as function of $R$ for different values of the
communication radius $r$. One can see that, for the same value of
$R$, $\langle T \rangle$ decreases as $r$ increases.

Figure~\ref{fig5} shows the load distribution $P(n)$ vs $n$ for
different values of $v$ and $r$, where $n$ represents the load and
$P(n)$ is the probability that an agent has the load $n$. We define
an agent's load $n_{i}$ as:
\begin{equation}
n_{i}=\frac{\sum n_{i}(t)}{\Delta T},
 \label{6}
\end{equation}
where $n_{i}(t)$ is the number of packets staying in $i$'s queue at
time $t$, $\Delta T$ is period of time (we set $\Delta
T=5\times10^{4}$) and the sum runs over a period of time $\Delta T$.
An agent's load $n_{i}$ reflects the average number of packets
staying in $i$'s queue at a time.

Figures~\ref{fig5}(a)-(d) show the load distribution $P(n)$ for
different values of the moving speed $v$ when the packet-generation
rate $R=100$ and the communication radius $r=1$. One can see that,
$P(n)$ approximately displays the Poisson distribution when $v$ is
very small (i.e., $v=0.001$ and $v=0.01$). As $v$ increases (i.e.,
$v=0.1$ and $v=0.8$), the load distribution becomes highly
homogeneous and all agents have almost the same load. The highest
load of an agent decreases as $v$ increases from $v=0.001$ to
$v=0.1$ [see Figs.~\ref{fig5}(a)-(c)]. It is noted that for the same
packet-generation rate, the average traveling time $\langle T
\rangle$ keeps almost unchanged for $v=0.001$ and $v=0.1$ [see
Fig.~\ref{fig4}(a)]. Thus, the change of the highest load for
$v=0.001$ and $v=0.1$ is caused by the alteration of network
structures. As the moving speed increases, each agent's neighbors
change more frequently and the topology of network turns from
quasistatic structure to dynamical structure. When $v$ increases
from 0.1 to 0.8, the average traveling time $\langle T\rangle$
increases and the load of an agent enhances [see Figs.~\ref{fig5}
(c) and (d)]. Since all agents have the same delivering ability $C$,
the transportation capacity of the whole network is determined by
the highest load of an agent. The increase of the highest load
indicates that the decrease of the network's transportation
capacity. Combining Fig.~\ref{fig4}(a) with Figs.~\ref{fig5}(a)-(d),
we can understand why the highest transportation capacity of
networks is reached at the moderate value of the moving speed.

Figures~\ref{fig5}(e)-(h) show the load distribution $P(n)$ for
different values of $r$ when $v$ is fixed to be 0.01. One can
observe that, $P(n)$ changes from the approximate Poisson
distribution to highly homogeneous distribution as the communication
radius $r$ increases. Besides, the highest load of an agent
decreases as $r$ increases. According to Fig.~\ref{fig4}(b), for the
same packet-generation rate, the average traveling time
 $\langle T \rangle$ decreases as $r$ increases. The decrease of $\langle T
 \rangle$ shortens the time steps that a packet stays in the system and relieves the traffic
 load, leading to the enhancement of the network's transportation
 capacity. Thus the phenomenon observed in Fig.~\ref{fig3}(b) can be
explained.

\section{Epidemic spreading}\label{sec:epidemic}

\begin{figure}
\begin{center}
\scalebox{0.4}[0.4]{\includegraphics{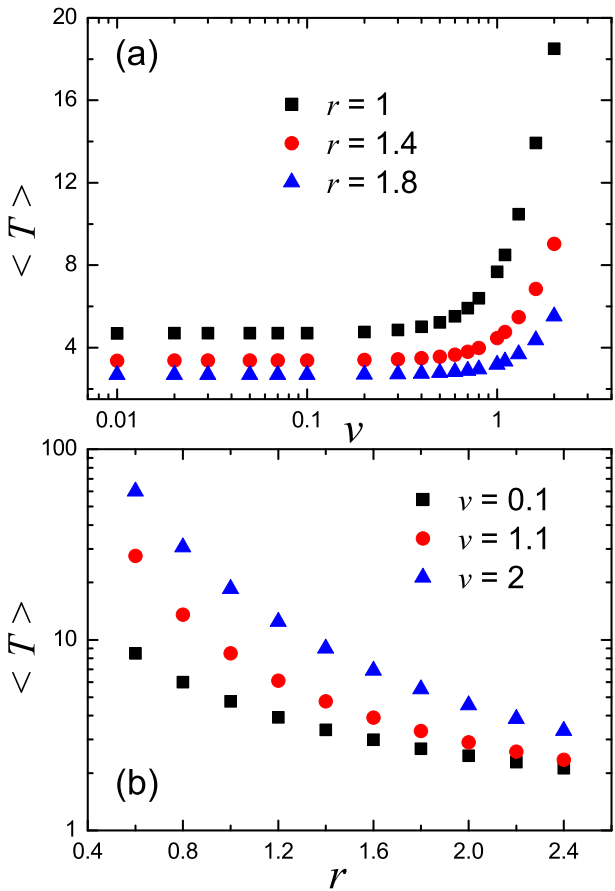}} \caption{(Color
online) (a) The average traveling time $\langle T \rangle$ as a
function of the moving speed $v$ for different values of the
communication radius $r$. (b) The dependence of $\langle T \rangle$
on $r$ for different values of $v$. }  \label{fig6}
\end{center}
\end{figure}

In this section, we study how the greedy routing affects the
traffic-driven epidemic spreading. We first consider the case where
each agent's delivering ability is infinite, $C\rightarrow \infty$,
so that traffic congestion will not occur in the network.

In the case of infinite delivering ability, the average traveling
time $\langle T \rangle$ is independent of packet-generation rate
$R$. Figure~\ref{fig6} shows the dependence of $\langle T \rangle$
on the moving speed $v$ and the communication radius $r$. For a
fixed value of $r$, $\langle T \rangle$ keeps almost unchanged for
$v<0.3$ but increases as $v$ continually increases [see
Fig.~\ref{fig6}(a)]. For a fixed value of $v$, $\langle T \rangle$
decreases as $r$ increases [see Fig.~\ref{fig6}(b)].

\begin{figure}
\begin{center}
\scalebox{0.4}[0.4]{\includegraphics{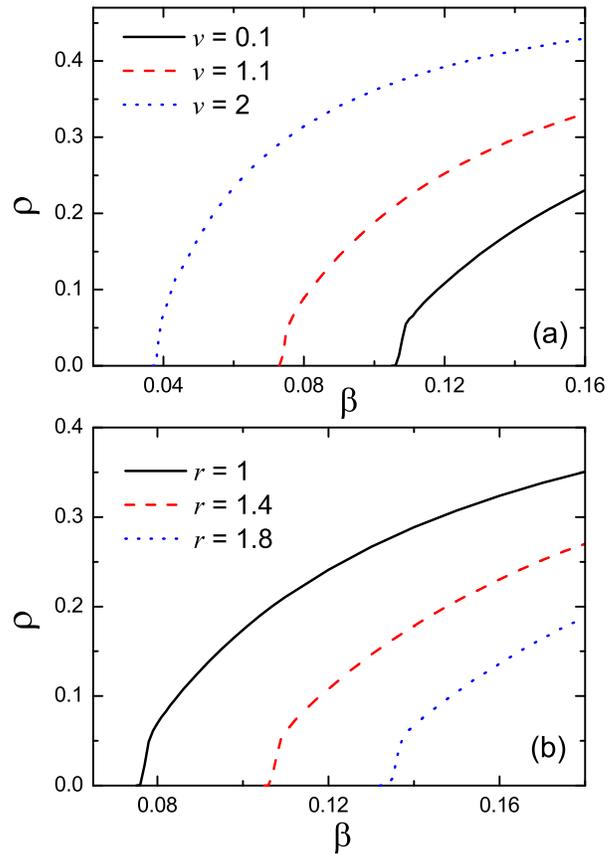}} \caption{(Color
online) (a) Density of infected agents $\rho$ as a function of the
spreading rate $\beta$ for different values of the moving speed $v$.
The communication radius $r=1.4$. (b) Density of infected agents
$\rho$ as a function of $\beta$ for different values of $r$. The
moving speed $v=0.1$. The packet-generation rate is $R=4000$ and
each agent's delivering ability is infinite. Each curve results from
an average over 50 different realizations. } \label{fig7}
\end{center}
\end{figure}

\begin{figure}
\begin{center}
\scalebox{0.4}[0.4]{\includegraphics{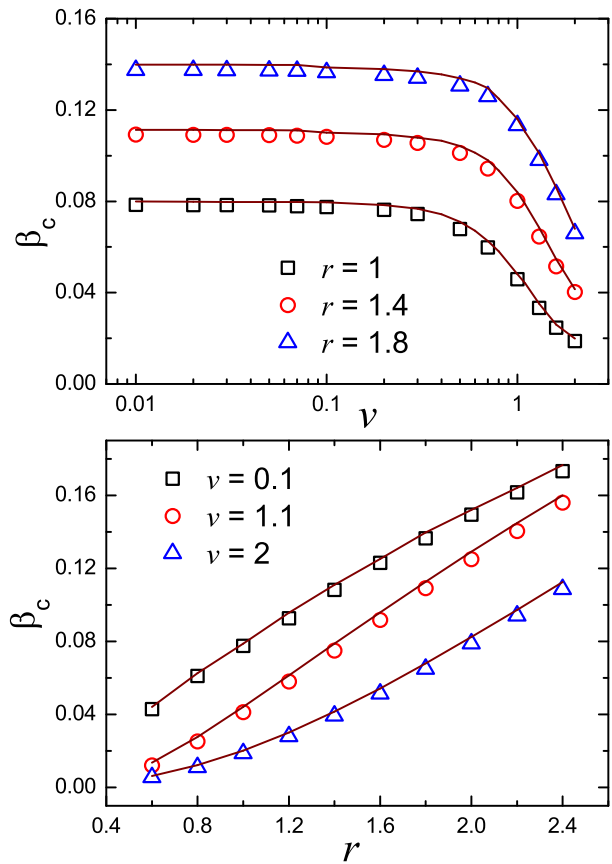}} \caption{(Color
online) (a) Epidemic threshold $\beta_{c}$ as a function of the
moving speed $v$ for different values of the communication radius
$r$. (b) Epidemic threshold $\beta_{c}$ vs $r$ for different values
of $v$. The packet-generation rate $R=4000$ and each agent's
delivering ability is infinite. Each data point results from an
average over 50 different realizations. The curves are theoretical
predictions from Eq. (\ref{9}). } \label{fig8}
\end{center}
\end{figure}

\begin{figure}
\begin{center}
\scalebox{0.4}[0.4]{\includegraphics{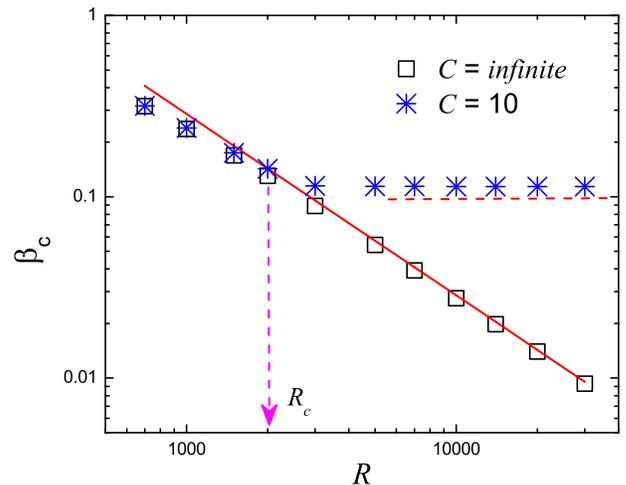}} \caption{(Color
online) Epidemic threshold $\beta_{c}$ as a function of the
packet-generation rate $R$ for $C=10$ and $C\rightarrow \infty$. The
moving speed $v=0.5$ and the communication radius $r=1$. For $C=10$,
the critical packet-generating rate $R_{c}$ is about 2000. Each data
point results from an average over 50 different realizations. The
solid and dash line is the theoretical prediction from Eq. (\ref{9})
and Eq. (\ref{11}), respectively. } \label{fig9}
\end{center}
\end{figure}

Figure~\ref{fig7} shows the density of infected agents $\rho$ as a
function of the spreading rate $\beta$ for different values of the
moving speed $v$ and the communication radius $r$. We observe that,
there exists an epidemic threshold $\beta_{c}$, beyond which the
density of infected agents $\rho$ is nonzero and increases as
$\beta$ is increased. For $\beta<\beta_{c}$, the epidemic dies out
and $\rho=0$. From Fig.~\ref{fig7}, one can find that both $v$ and
$r$ can affect the density of infected agents and the epidemic
threshold. Large values of $v$ and small values of $r$ promote the
epidemic spreading.

Figure~\ref{fig8}(a) shows the epidemic threshold $\beta_{c}$ as a
function of the moving speed $v$ for different values of the
communication radius $r$. One can see that, for a fixed value of
$r$, $\beta_{c}$ keeps almost unchanged for $v<0.3$ but decreases as
$v$ continually increases. Figure~\ref{fig8}(b) shows the dependence
of $\beta_c$ on $r$ for different values of $v$. One can observe
that $\beta_c$ increases as $r$ increases. In the following, we will
provide a theoretical analysis to calculate the epidemic threshold
$\beta_c$.

Since all agents have the same communication radius $r$ and the load
distribution $P(n)$ follows the approximate Poisson distribution or
highly homogeneous distribution, the mean-field theory can be
applied. At each time step, the average number of packets that an
agent delivers is $R\langle T \rangle/N$. Thus the rate equation for
the epidemic dynamics can be written as
\begin{equation}
\frac{d\rho(t)}{dt}=-\rho(t)+\frac{R\langle T \rangle}{N}
\beta\rho(t)[1-\rho(t)].
 \label{7}
\end{equation}
After imposing the stationarity condition $d\rho(t)/dt = 0$, we
obtain
\begin{equation}
\rho[-1+\frac{R\langle T \rangle}{N}\beta(1-\rho)]=0.
 \label{8}
 \end{equation}
From Eq.(\ref{8}), we get the epidemic threshold
\begin{equation}
\beta_{c}=\frac{N}{R\langle T \rangle}.
 \label{9}
 \end{equation}

The comparison between numerical and theoretical values of
$\beta_{c}$ is shown in Fig.~\ref{fig8} (we use the numerical values
of $\langle T \rangle$ in Eq. (\ref{9}) because of the difficulty in
calculating $\langle T \rangle$ theoretically). From
Fig.~\ref{fig8}, we find that the theoretical prediction of
$\beta_{c}$ is in good agreement with that of simulation result.

Next, we consider the finite delivery ability. In this case, traffic
congestion occurs when the packet-generation rate $R$ is above a
critical value $R_{c}$. Figure~\ref{fig9} shows the epidemic
threshold $\beta_c$ as a function of $R$ for $C=10$ and
$C\rightarrow \infty$. One can see that, $\beta_c$ scales inversely
with the $R$ for $C\rightarrow \infty$, as predicted by Eq.
(\ref{9}). For $C=10$, $\beta_c$ decreases to a steady value as $R$
increases. When $R \leq R_c$, $\beta_{c}$ is almost the same for
finite and infinite delivery ability. However, for $R> R_c$, we
observe that the value of $\beta_c$ is larger for $C=10$ than that
of $C\rightarrow \infty$, indicating that traffic congestion can
suppress the spreading of disease. Similar phenomena were also been
found in Refs.~\cite{PNAS,routing2}.

In the case of finite delivery ability, the lower limit of the
epidemic threshold can be estimated as follows.

For sufficiently large values of $R$, all agents will become
congested and each agent can deliver only $C$ packets at each time
step. Thus Eq. (\ref{7}) must be revised as

\begin{equation}
\frac{d\rho(t)}{dt}=-\rho(t)+C \beta\rho(t)[1-\rho(t)].
 \label{10}
\end{equation}
Imposing the stationarity condition $d\rho(t)/dt = 0$ and we obtain
the epidemic threshold
\begin{equation}
\beta_{c}=\frac{1}{C}.
 \label{11}
 \end{equation}

\section{Conclusion} \label{sec:conclusion}

In conclusion, we studied how greedy routing affects traffic
dynamics and traffic-driven epidemic spreading on networks of mobile
agents. Our main findings are the following. Firstly, for finite
delivering ability, the transportation capacity of networks peaks at
the moderate moving speed and increases as the communication radius
increases. Secondly, the average traveling time of a packet
increases as the moving speed increases but decreases as the
communication radius increases. Thirdly, for infinite delivering
ability, the epidemic threshold decreases as the moving speed
increases and increases as the communication radius increases.
Fourthly, traffic congestion can suppress epidemic spreading. Since
the study of networks of mobile agents has received increasing
attention in recent years, our results are valuable for
understanding the structure and dynamics of mobile networks.

\begin{acknowledgments}

HXY and BHW were funded by by the National Important Research
Project (Grant No. 91024026), the National Natural Science
Foundation of China (No. 10975126), and the Specialized Research
Fund for the Doctoral Program of Higher Education of China (No.
20093402110032).  WXW and YCL were supported by AFOSR under Grant
No. FA9550-10-1-0083 and by NSF under Grants No. BECS-1023101 and
No. CDI-1026710.

\end{acknowledgments}

\end{document}